\documentclass[10pt]{iopart}
\usepackage{amssymb}
\usepackage{graphicx}
\usepackage{enumerate}
\begin{document}
\title{A spin-dependent local moment approach to the Anderson impurity model}
\author{Choong H Kim$^1$ and Jaejun Yu$^{1,2}$}
%\email[Corresponding author.]{Email : jyu@snu.ac.kr}
\address{$^1$ Department of Physics $\&$ Astronomy and Center for Strongly
  Correlated Materials Research, Seoul National University, Seoul 151-747, Korea} 
\address{$^2$ Center for Theoretical Physics, Seoul National University, Seoul 151-747, Korea} 
\ead{jyu@snu.ac.kr}
%\tolerance=50
\begin{abstract}
  We present an extension of the local moment approach to the Anderson
  impurity model with spin-dependent hybridization.  By employing the two
  self-energy description, as originally proposed by Logan and co-workers,
  we applied the symmetry restoration condition for the case with
  spin-dependent hybridization.  Self-consistent ground states were
  determined through the variational minimization of ground state
  energy. The results obtained with of our spin-dependent local moment approach applied to a
  quantum dot system coupled to ferromagnetic leads are in good
  agreement with those obtained from previous work using numerical renormalization group
  calculations.
\end{abstract}  

%\submitto{\JPCM}

\section{Introduction}

The Anderson impurity model (AIM) \cite{anderson} and its extensions have been a matter of
central importance in the recent developments of condensed matter physics.
The AIM serves not only as a prototype model for the Kondo effect \cite{thebook}
but also as a key ingredient in the dynamical mean-field theory (DMFT) \cite{dmft1,dmft2}
for  strongly correlated electron systems. 
Recent advances in quantum dot (QD) experiments have demonstrated that a quantum dot
connected to leads can act as a magnetic impurity in metal so that the
Kondo-type behavior emerges at low temperatures
\cite{kouwenhoven}. Indeed the Kondo effect in quantum dot systems have
been probed by many theoretical \cite{glazman,ng} and experimental studies
\cite{goldhaber,cronenwett,vanderwiel}.
The fine tunability of control parameters in quantum dot systems has
spurred the investigation of Kondo physics in various aspects including
a quantum dot coupled to ferromagnetic leads.  A flood of very
recent works
\cite{bulka,lopez,mschoi,martinek1,martinek2,gazza}
has focused on this issue as motivated by its potential
applications to spintronics.

Although there are many theoretical approaches to the solution of the AIM,
it is still not easy to deal with the AIM coupled to ferromagnetic leads,
where spin-dependent charge fluctuations need to be taken into account.
In other words, it is necessary to get a non-perturbative method which can deal with
both charge and spin excitation channels.
For instance, some of slave-boson mean-field calculations could not
describe the finite splitting of the Kondo peak \cite{bulka,lopez}
in the spin-dependent case due to the absence of charge fluctuations.
Although the numerical renormalization group (NRG) calculation
is known to provide accurate results for the impurity problem,
the standard NRG technique seems not good enough for the spin-polarized system
due to the absence of energy scale separation \cite{hofstetter}.

As a way to find an effective approach to the quantum dot coupled to
ferromagnetic leads, we have considered the local moment approach (LMA)
originally suggested by Logan and co-workers \cite{symLMA,LMA}.  Based on
the intuitive notion of local moment fluctuations and the
symmetry restoration condition, this approach is
technically simple and transparent and yet can cover all energy
scales and interaction strengths \cite{NRGvsLMA}.
Since it relies on the degeneracy of the two mean-field saddle point configurations,
however, it is not obvious how to handle the case with spin polarization.
The original LMA of the Anderson impurity model in an external
field \cite{sLMA1,sLMA2}
did not treat each spin component of the full Green's function separately,
but just calculated the sum.
In this work,
we extended the LMA formalism to include the spin-dependent hybridization.
The basic idea is a generalization of the two-self-energy description,
allowing the variation of the weights of the spin-up and
spin-down configurations. The weights of each local moment configuration
can be determined through the minimization of the total energy of impurity system.

This paper is organized as follows.  In section~\ref{sec:Model}, we give a
general description of model Hamiltonian for the quantum dot coupled
to ferromagnetic leads. In section~\ref{sec:Formalism}, we describe the
formalism of our spin-dependent LMA (to be abbreviated as sLMA) including basic ideas on the
generalization of the two-self-energy description, the symmetry restoration
condition, and the determination of configuration weights. The calculated
spectral function for the asymmetric Anderson model is given in
section~\ref{sec:Result}, together with a comparison with previous works.

\section{Model}
\label{sec:Model}

\begin{figure}
\centering
\includegraphics[width=0.5\textwidth]{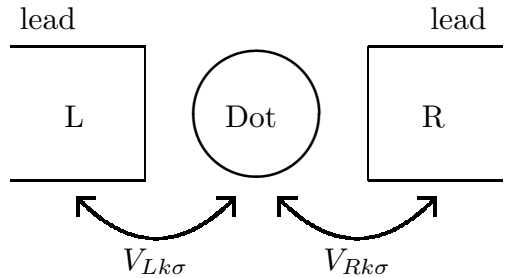}
\caption{Quantum dot coupled to ferromagnetic leads.}
\label{fig:qd}
\end{figure}

Let us consider a model for the quantum dot (QD) coupled to two ferromagnetic leads.
Figure \ref{fig:qd} shows a schematic view of the QD system coupled to
ferromagnetic leads with different spin polarization directions.
We assume that a single level QD with a charging energy $U$.
It is well known that the QD is equivalent to a single-level impurity state \cite{glazman,ng} in the AIM.
In this system, two cases with parallel (P) and antiparallel (AP) magnetic configurations are possible
for the ferromagnetically ordered leads.
By making the canonical transformation \cite{glazman,mschoi},
both P and AP configurations could be mapped onto an effective model with a single lead.
After the transformation, the AP configuration was shown to be
equivalent to a usual QD coupled to a single non-magnetic lead.
Thus, from now on, we will focus on the P configuration
where the spin-dependence becomes explicit.
With the Fermi energy kept at the energy origin ($\epsilon_F=0$),
the Hamiltonian for the P configuration can be represented by
\begin{eqnarray}\label{modelH}
  \mathcal{H} = \sum_\sigma (\epsilon_i -\sigma h) d^\dag_\sigma d_\sigma + U
  d^\dag_\uparrow d_\uparrow d^\dag_\downarrow d_\downarrow
  + \sum_{k \sigma} \left[ \epsilon_{k\sigma} c^\dag_{k \sigma} c_{k \sigma} 
    + V_{k\sigma} d^\dag_\sigma c_{k \sigma} + V^*_{k\sigma} c^\dag_{k \sigma} d_\sigma \right] 
\end{eqnarray}
where $c^\dag_{ k \sigma}$ $(c_{ k \sigma})$ is the creation
(annihilation) operator for an electron with the wave vector $k$ and
spin $\sigma$ in the leads.  And $d^\dag_\sigma$ $(d_\sigma)$ is the
creation (annihilation) operator for electrons in the dot.
The local Zeeman coupling energy $h=\frac12 g\mu_B H$ is included for the sake of generality.
Here it is noted that the effective dot-lead coupling $V_{k\sigma}$
is connected to the original QD model in figure \ref{fig:qd}:
\begin{equation}
V_{k\sigma} = \frac1{\sqrt2} ( V_{Lk\sigma} + V_{Rk\sigma} ).
\end{equation}
Considering its parametric dependence, the spin-dependent tunneling amplitude $V_{\sigma k}$ can be described by
introducing a polarization parameter $p$ for the spin-dependent host density-of-states (DOS) $\rho_\sigma(\omega)$:
\begin{equation}
  \label{eq:DOS}
  \rho_{\sigma}(\omega) = \rho (1+ \sigma p)~~~~~~~~{\rm for}~-D<\omega<D.
\end{equation}
Neglecting the dependence on $k$ of $V_{\sigma k}$ in the large band width limit of conduction electrons,
the spin-dependent hybridization parameter $\Delta_\sigma = \sum_k |V_k|^2(\omega^+ - \epsilon_{k\sigma})^{-1}$
can be approximated by a simplified form:
\begin{equation}
\Delta_{\sigma} (\omega) \simeq - {\rm i} \Delta_{0}(1+\sigma p){\mathrm{sgn}(\omega)}
\end{equation}
where $\Delta_0 = \pi V^2 \rho$.
Some authors have pointed out that the shape of the conduction band
may be important in this problem \cite{martinek2,gazza}.
However, for the sake of simplicity, we only deal with the $D\rightarrow\infty$ limit
without considering the detailed band structure.
Therefore, the Green's function of the d electrons can take a simple form of
\begin{equation}
g^{-1}_{\sigma}(\omega) = \omega^+ - \epsilon_{i} - \sigma h  + \rm{ i} \Delta_0(1+\sigma p){\mathrm{sgn}(\omega)}.
\end{equation}
in the non-interacting limit ($U=0$).
In what follows, we consider the case of $h=0$ only.

\section{Formalism}
\label{sec:Formalism}
\subsection{Local moment approach}

The LMA starts off based on the mean-field, i.e., unrestricted-Hartree-Fock (UHF), solutions of the AIM.
For large $U$ values,
the UHF treatment of the single-orbital AIM gives the doubly degenerate local moment solutions \cite{anderson}.
Two solutions are denoted by $\alpha=A$ or $B$, corresponding to the local moments $\mu=+|\mu|$ and $-|\mu|$, respectively.
In order to remedy the broken symmetry nature of the UHF solutions,
LMA employs the two-self-energy description (TSE):
\begin{equation}
  \label{eq:TSE}
G_\sigma(\omega) = \frac12 [ G_{A\sigma}(\omega) +  G_{B\sigma}  (\omega) ].
\end{equation}
In this framework, the  LMA is designed to describe
the  local moment at the impurity site, fluctuating between the two configurations, $A$ and $B$.
In addition, to include the dynamics of the self-energy going beyond the static UHF self-energy,
the LMA uses the HF Green's function as a bare propagator for the calculations of the higher-order diagrams
(which will be discussed in detail in section~\ref{subsec:self-energies}).
The higher-order terms for the self-energy are indeed enough for
describing the Fermi liquid behavior at low energy scale, but to do this
it is required to introduce the symmetry restoration (SR) condition and the Friedel sum rule
for recovery of the correct low energy behavior.
The SR condition plays a role in linking the two configurations $A$ and $B$ at low energies.
Taking advantage of its efficiency and transparency,
the LMA has been applied to the lattice model
within the framework of dynamical mean-field theory \cite{LMAdmft1,LMAdmft2,LMAdmft3,LMAdmft4}.

\subsection{Mean-field approximation}
In the absence of spin polarization, the UHF
approximation gives two degenerate broken symmetry solutions:
$\mu=+|\mu|$ and $-|\mu|$.
However, in the presence of spin polarization, 
the two mean-field saddle points become no more degenerate.
To distinguish two solutions, we labeled two configurations by $\alpha=A$ or $B$,
corresponding to the local moment $\mu=+|\mu_A|$ and $-|\mu_B|$, respectively.

Taking account of the configurations $A$ and $B$, we obtained the
mean-field propagator $\mathcal{G}_{\alpha\sigma}$ for $\alpha=A,B$ and $\sigma=\uparrow,\downarrow$:
\begin{eqnarray}
\mathcal{G}^{-1}_{A\sigma} (\omega) &= \omega^+ - e_{iA} +\sigma x_A + {\rm i} \Delta_0 (1+\sigma p) {\mathrm{sgn}(\omega)}\nonumber\\
\mathcal{G}^{-1}_{B\sigma} (\omega) &= \omega^+ - e_{iB} -\sigma x_B + {\rm i} \Delta_0 (1+\sigma p) {\mathrm{sgn}(\omega)}
\end{eqnarray}
where
\begin{equation}
x_\alpha = \frac12 U |\mu_\alpha|,~~~~
e_{i\alpha} = \epsilon_{i\alpha} + \frac12 U n_\alpha,
\end{equation}
Here we introduced the configuration-dependent on-site energy, $\epsilon_{i\alpha}$,
which may be necessary for describing the $\alpha$-dependent renormalization
of the on-site energy under the asymmetric configuration.
And the corresponding spectral densities $D^0_{\alpha\sigma}(\omega)$ are
\begin{eqnarray}
D^0_{\alpha\sigma} (\omega) &= -\frac{1}{\pi}{\mathrm{sgn}(\omega)} \mathrm{Im} \mathcal{G}_{\alpha\sigma} (\omega).
\end{eqnarray}
Consequently, the mean-field charge and moment could be determined  from the self-consistent solution of
the following equations:
\begin{eqnarray}
{\bar{n}}_{\alpha\sigma} &= \int^0_{-\infty} d\omega D^0_{\alpha\sigma}(\omega;e_{i\alpha},x_\alpha) \nonumber\\
{\bar{n}}_\alpha &= \int^0_{-\infty} d\omega \Big[ D^0_{\alpha\uparrow}(\omega;e_{i\alpha},x_\alpha) + D^0_{\alpha\downarrow}(\omega;e_{i\alpha},x_\alpha)  \Big]\nonumber\\
|{\bar{\mu}}_\alpha| &= \alpha \int^0_{-\infty} d\omega \Big[ D^0_{\alpha\uparrow}(\omega;e_{i\alpha},x_\alpha) - D^0_{\alpha\downarrow}(\omega;e_{i\alpha},x_\alpha)  \Big],
\end{eqnarray}
where  $\alpha=+$ and $-$ for $A$ and $B$, respectively.

\subsection{Two-self-energy description}
For the spin-dependent case, where the up-spin and down-spin symmetry no longer exists,
we generalized the LMA description of the Green's function
by allowing the two saddle point configurations to have different weights.
Extending the original two-self-energy description,
the average Green's function could be expressed as
\begin{equation}\label{eq:tse}
G_\sigma(\omega) = c_A G_{A\sigma}(\omega) + (1-c_A) G_{B\sigma}  (\omega),
\end{equation}
where the Green's function for each configurations $\alpha$ is provided as
\begin{equation}
G^{-1}_{\alpha\sigma}(\omega) = g^{-1}_{\alpha\sigma}(\omega) - {\tilde{\Sigma}}_{\alpha\sigma}(\omega)
\end{equation}
with the $\alpha$-dependent $g_{\alpha\sigma}$ and the corresponding self-energy ${\tilde{\Sigma}}_{\alpha\sigma}(\omega)$:
\begin{equation}
g^{-1}_{\alpha\sigma}(\omega) = \omega^+ -{\epsilon_{i\alpha}} + {\rm i} \Delta_0(1+\sigma p){\mathrm{sgn}(\omega)} .
\end{equation}
The configuration weight $c_A$ or $(1-c_A)$ reflects the probability of $A$ or $B$ configuration.
In the absence of spin-dependence, $c_A$ is trivially equal to $\frac12$.
The dependence on $\alpha$ of $g_{\alpha\sigma}(\omega)$ arises from the relative
chemical potential shift of the $A$ and $B$ configurations due to the
$\alpha$-dependent renormalization of the bare energy level.

\subsection{Self-energies}
\label{subsec:self-energies}

\begin{figure}
\begin{center}
\includegraphics[width=0.5\textwidth]{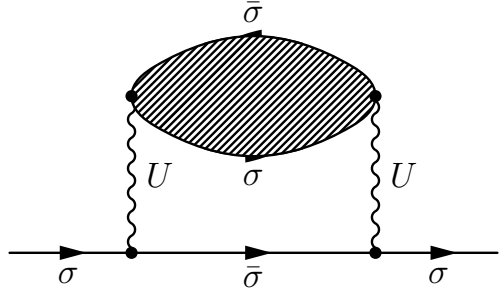}
\end{center}
\caption{Principal contribution to the LMA $\Sigma_{\sigma}(\omega)$.
The shaded bubble represents the polarization propagator as
given by the RPA-like particle-hole ladder sum in the transverse spin
channel.
}
\label{fig:self}
\end{figure}

The evaluation of self-energy in the spin-dependent LMA (sLMA) followed the same steps
prescribed in the standard LMA procedure,
the details of which are given in \cite{symLMA}.
The difference in our procedure lies mainly on
the spin-dependent terms due to the non-degenerate local moment configurations.

Following the notations in \cite{symLMA}, the self-energies were divided into the static HF term and the rest:
\begin{eqnarray}
{\tilde{\Sigma}}_{\alpha\sigma}(\omega) &= {\tilde{\Sigma}}^0_{\alpha\sigma} + \Sigma_{\alpha\sigma}(\omega) \nonumber\\
&= \frac{U}2({\bar{n}}_\alpha-\alpha\sigma|{\bar{\mu}}_\alpha|) + \Sigma_{\alpha\sigma}(\omega).
\end{eqnarray}
The purely static contribution ${\tilde{\Sigma}}^0_{\alpha\sigma}$ is given by the UHF calculation.
And all of the dynamics is contained in the
$\Sigma_{\alpha\sigma} (\omega) = \Sigma_{\alpha\sigma} [\{\mathcal{G}_{\alpha\sigma}\}]$,
which is a functional of the underlying MF propagator $\mathcal{G}_{\alpha\sigma} (\omega)$,
as shown in figure \ref{fig:self}:
\begin{eqnarray}\label{dyself}
\Sigma_{\alpha\sigma}(\omega) = U^2 \int^\infty_{-\infty} \frac{d\omega_1}{2\pi i} \mathcal{G}_{\alpha\bar{\sigma}} (\omega+\omega_1) \Pi^{\sigma\bar{\sigma}}_{\alpha\alpha} (\omega_1).
\end{eqnarray}
This diagram contains the dynamical spin-flip processes,
and within the random phase approximation (RPA) the polarization diagrams are calculated,
\begin{equation}\label{RPA}
\Pi^{\sigma\bar{\sigma} }_{\alpha\alpha}(\omega) = \frac{{\ ^0 \Pi}^{\sigma\bar{\sigma} }_{\alpha\alpha}(\omega)}{1- U {\ ^0 \Pi}^{\sigma\bar{\sigma} }_{\alpha\alpha}(\omega)}
\end{equation}
with the bare particle-hole bubble
\begin{equation}\label{barebubble}
{\ ^0 \Pi}^{\sigma\bar{\sigma} }_{\alpha\alpha}(\omega) = {\rm  i} \int^\infty_{-\infty} \frac{d\omega_1}{2\pi} \mathcal{G}_{\alpha\bar{\sigma}} (\omega_1) {\mathcal{G}}_{\alpha\sigma} (\omega_1-\omega).
\end{equation}

\subsection{Symmetry Restoration}

The symmetry restoration (SR) is required to recover the Fermi liquid behavior at low $\omega$.
To achieve the same goal in the spin-dependent formalism, we have to modify the original SR condition for
the generalized two-self-energy description with 
the different weights $c_A$ and $(1-c_A)$ for spin-up and down configurations, respectively.
In terms of self-energies, we can rewrite (\ref{eq:tse}) as
\begin{equation}
\frac1{ g^{-1}_\sigma - {\Sigma}_\sigma} = \frac{c_A}{  g^{-1}_{A\sigma} - {\tilde{\Sigma}}_{A\sigma} } + \frac{1-c_A}{  g^{-1}_{B\sigma} - {\tilde{\Sigma}}_{B\sigma} }.
\end{equation}
We could consider the following relation between the single self-energy $\Sigma_\sigma(\omega)$ and the two-self-energies
$\Sigma_{\alpha\sigma}(\omega)$:
\begin{eqnarray}
\epsilon_i + \Sigma_\sigma(\omega) &= c_A f_{A\sigma}(\omega) + (1-c_A)f_{B\sigma}(\omega)\nonumber\\
&+\frac{ c_A(1-c_A)[f_{A\sigma}(\omega) - f_{B\sigma}(\omega )]^2}{\omega^+ + {\rm i}\Delta_0(1+\sigma p){\rm sgn}(\omega) - c_A f_{B\sigma}(\omega) - (1-c_A)f_{A\sigma}(\omega)  },
\end{eqnarray}
where $f_{\alpha\sigma}(\omega) = \epsilon_{i\alpha} +\tilde{\Sigma}_{\alpha\sigma}(\omega)$ is defined with $\alpha=A,B$.
Since the two-self-energies vanish at the Fermi level, i.e., ${\rm Im} \Sigma_{\alpha\sigma}(\omega=0)=0$,
by imposing the condition that the imaginary part of self-energy at the Fermi level must be equal to zero,
we could obtain the following generalized SR condition:
\begin{equation}\label{GSR}
\epsilon_{iA} + {\tilde{\Sigma}}^{{\rm{R}}}_{A\sigma}(\omega=0) = \epsilon_{iB} + {\tilde{\Sigma}}^{{\rm{R}}}_{B\sigma}(\omega=0).
\end{equation}
This SR condition contains the relative chemical potential shift of two configurations.  
One can readily note that this generalized version of
SR reduces to the original SR condition when there is no relative chemical
potential shift, i.e.,  $\epsilon_{iA}=\epsilon_{iB}$.
In addition, when the SR condition of (\ref{GSR}) is satisfied,
the Friedel sum rule \cite{sumrule,luttinger} equation can be written as
\begin{equation}\label{sumrule3}
\epsilon_{i\alpha} + U{\bar{n}}_{\alpha\bar{\sigma}} + \Sigma^{{\rm{R}}}_{\alpha\sigma} (\omega=0;e_{i\alpha},x_\alpha)
=(1+\sigma p) \Delta_0 \tan\left[\frac{\pi}2 (1-2n^\sigma_{\rm{imp}})\right].
\end{equation}
This is complementary to the previous physical arguments
related to the Fermi liquid behavior at the Fermi level.
Here our self-energy arguments assert that the SR must be taken
into account, when we are dealing with the two-self-energy descriptions.

\subsection{Filling constraint}
\label{subsec:filling}
To make the practical calculation feasible,
we introduced a physically motivated approximation.
that the total impurity charge $n_{\rm{imp}}$ does not depend on $h$ or $p$.

\begin{equation}\label{fillconstraint}
n_{\rm{imp}}(h=0;p=0) = n_{\rm{imp}}(h,p) = n^A_{\rm{imp}} (h,p) = n^B_{\rm{imp}} (h,p)
\end{equation}
where
\begin{eqnarray}
n^\alpha_{\rm{imp}} = \int_{-\infty}^0 d\omega \Big[ D_{\alpha\uparrow}(\omega) + D_{\alpha\downarrow} (\omega) \Big].
\end{eqnarray}
Within this assumption,
we fixed the value of $n_{\rm{imp}}$ as calculated for the $h=p=0$ system.
Thus, we were able to determine $\epsilon_{iA}$ and $\epsilon_{iB}$ for a given $n_{\rm{imp}}$.

Although the best way to do this is the self-consistent determination of $n^\alpha_{\rm{imp}}$,
there is no practical method available to determine the $n^\alpha_{\rm{imp}}$ self-consistently at present.
Since the assumption on the independence of $n_{\rm{imp}}$ on $h$ and $p$ is subtle
but important in the real calculations,
it may deserve more explanation.
At least, both the exact Bethe ansatz solutions \cite{tsvelik} and the NRG calculation \cite{hofstetter}
show the independence of $n_{\rm{imp}}$ on $h$.
Indeed this result supports that our assumption is valid at least for the $h\neq0$ case.
Even though the there is no explicit report for the case of $p\neq0$, 
we can say that this assumptions is a good approximation for two cases
where the charge fluctuations are strongly suppressed by either the particle-hole symmetry
or the large on-site interaction, i.e., $U\gg\Delta_0$.

\subsection{Variational Principle}

\begin{figure}%[h]
\begin{center}
\includegraphics[width=0.4\textwidth]{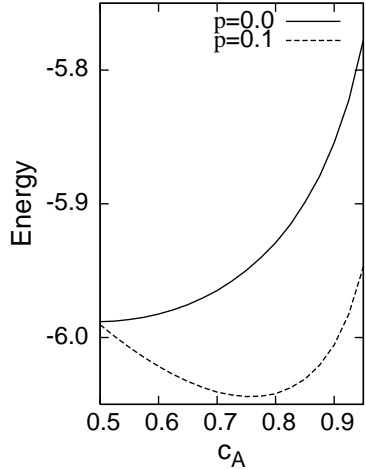}
\end{center}
\caption{Energy vs. $c_A$ curve for ($U/\Delta_0=12.5$, $\epsilon_i/\Delta_0=-5$ and $p=0.0,0.1$). 
The other configurations (e.g. $p=0.3,0.5$) has quite different energy
scale to display in one figure.}
\label{fig:CvsE}
\end{figure}

For the ground state energy of the system, we adopted the following expression \cite{totalE}:
\begin{equation}
E = \sum_\sigma \int_{-\infty}^0 \frac1\pi \mathrm{Im}\left\{ \left[\omega-\frac12 {\tilde{\Sigma}}_\sigma(\omega) \right] G_\sigma(\omega) \right\} d \omega,
\end{equation}
which is certainly valid only for the wide-band limit.
It may be necessary to include the contribution from the conduction electron part for the general case.
In the spin-dependent LMA, $c_A$ was taken as a variational parameter for the energy minimization.
For a given $c_A$,
there exists a corresponding set of LMA parameters ($\epsilon_{iA},\epsilon_{iB},e_{iA},e_{iB},x_A,x_B$)
which satisfies the self-consistent equations (SR, Friedel sumrule and filling constraint).
Hence we were able to calculate the ground state energy for the given $c_A$ which minimizes the total energy.

Figure \ref{fig:CvsE} shows the energy versus $c_A$ curve for $U/\Delta_0=12.5$ and $\epsilon_i/\Delta_0=-5$.
(Here we display the results of $c_A>0.5$ considering the symmetry.)
For the $p=0$ case, the minimum is present at $c_A=0.5$ as expected.
It signifies that our energy variation scheme works for $p=0$.
For the case of $p=0.1$, the energy minimum is at $c_A\simeq0.77$.

\section{Results}
\label{sec:Result}

\begin{figure}%[h]
\centering
\includegraphics[width=0.7\textwidth]{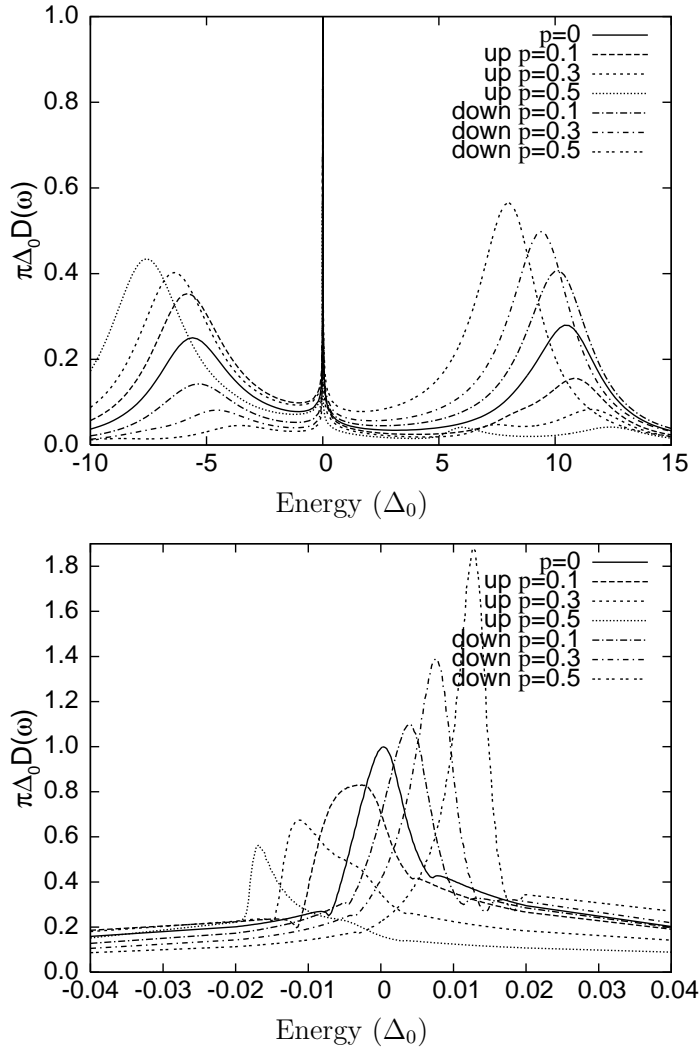}\\
\caption{Local DOS of the QD for the asymmetric Anderson model
  ($U/\Delta_0=15$ and $\epsilon_i/\Delta_0=-5$). Note  the different energy scales in two figures.}
\label{fig:spectral}
\end{figure}

\begin{figure}%[h]
\centering
\includegraphics[width=0.5\textwidth]{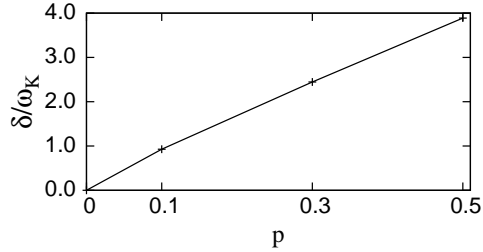}
\caption{Splitting $\delta$ of the Kondo peak as a function of $p$ for $U/\Delta_0=15$ and $\epsilon_i/\Delta_0=-5$.
$\omega_K$ is defined as the FWHM of the $p=0$ Kondo peak.}
\label{fig:dvsp}
\end{figure}

We now turn to the sLMA calculation results.
Figure \ref{fig:spectral} shows the spectral functions 
$\pi\Delta_0D_\uparrow(\omega)$ and $\pi\Delta_0D_\downarrow(\omega)$,
i.e., local DOS of the QD, for the different values of lead polarization $p$ in the P configuration.
Please note the different energy scales in two figures.
Our results on the finite splitting of Kondo peak are found to be consistent with those of
numerical renormalization group (NRG) calculations \cite{mschoi,martinek2}.
As $p$ increases, the Kondo peaks for $\pi\Delta_0D_\uparrow(\omega)$ and $\pi\Delta_0D_\downarrow(\omega)$
shift in the opposite directions and the Kondo peaks split into two
and the values of the spectral functions at Fermi level decrease.
As a result, Kondo effects are reduced or suppressed in the presence of ferromagnetic leads.
The relatively small magnitudes of the splitting are attributed to the similar underestimation of
the width of Kondo peaks in the standard LMA. 
As shown in figure \ref{fig:dvsp}, qualitative features such as dependence on $p$ of Kondo peak splitting $\delta$ are
in good agreement with NRG.
One can observe that $\delta$ is linear in the $p$, confirming the NRG result \cite{mschoi}.

The finite splitting of Kondo peaks may be understood via Haldane's scaling arguments:
the renormalization of the on-site energy is spin-dependent due to the spin-dependence of the hybridization;
\begin{equation}
\Delta \epsilon \simeq \frac{p\Delta_0}{\pi} \ln \left( \frac{ |\epsilon_i| }{|U+\epsilon_i|} \right),
\end{equation}
where $\Delta \epsilon$ was attributed to the splitting of the renormalized levels \cite{martinek2}.
Then, the coupling acts as an effective magnetic field, leading to the finite splitting.

In figure \ref{fig:spectral}, one can note
that the mean-field peaks are also shifted in opposite directions.
In the LMA scheme, the shift of the  mean-field peak, arises from the different
chemical potential shift of $A$ and $B$, i.e., $\epsilon_{iA} \neq
\epsilon_{iB}$.  When we fixed the $\epsilon_{iA} =
\epsilon_{iB} = \epsilon_{i}$, we could not observe the shifts
although the weights of peaks become drastically different.
The Hubbard satellites were more pronounced in sLMA,
while the same mean-field peak shifts are rather small in the NRG calculations.
It is well known that the high energy features are usually underrated in the calculation of 
dynamical properties with the NRG method \cite{bulla}.
In addition, this problem becomes more serious in the spin symmetry breaking system
due to the absence of energy scale separation \cite{hofstetter}.

In fact, the Kondo peak of the down-spin electron in our results is remarkable to observable.
The peaks of down-spin is higher than the peaks of up-spin with the same $p$.
Even though this is consistent with NRG results, there are some problems.
In the NRG method, the height of the Kondo peak is decreased as increase of $p$,
but our results show the opposite trend.
The filling constraint which is introduced as the physically motivated approximation in section \ref{subsec:filling}
could be the possible source of this difference.
But it is very difficult to work out the source of the problem at this stage.
Treating this problem could be an important issue for future work.

\section{Conclusion}\label{chap:conclusion}
In this paper we introduced a spin-dependent local moment approach (sLMA) for the study of the Anderson impurity model with spin-dependent hybridization.
As an extension of the standard local moment approach to the spin-dependent system,
we employed the generalized two-self-energy description and symmetry restoration to deal with spin polarizations.
The approach has been applied to a quantum dot system which is coupled to ferromagnetic leads.
Our results for the asymmetric Anderson model are in a qualitative agreement with the NRG results.

\ack
We would like to thank Professor G. S. Jeon for helpful discussions.
This work was supported by the KOSEF through CSCMR SRC and by the KRF Grant (MOEHRD KRF-2005-070-C00041).

\end{document}